\expandafter\ifx\csname natexlab\endcsname\relax\fi

\documentclass[numberedappendix]{emulateapj}
\slugcomment{{\sc Accepted by ApJ:} November 11th, 2019.} 
\usepackage{graphicx}
\usepackage{epstopdf}
\usepackage{color}
\usepackage{amsmath}
\usepackage{mathtools}
\usepackage{hyperref}
\usepackage{perpage}
\MakePerPage{footnote}
\def\a4{\hsize 17.0cm \vsize 25.cm}

\defcitealias{MRN1977}{MRN} 
\defcitealias{MartinezGonzalezetal2016}{MTS16}
\defcitealias{MartinezGonzalezetal2017}{MWP17}
\defcitealias{MartinezGonzalezetal2018}{MWP+18}
\defcitealias{Kroupa2001}{Kroupa}
\defcitealias{Bocchioetal2016}{B16}
\defcitealias{Ferraraetal2016}{Ferrara et al.}
\shorttitle{}

\shorttitle{Dusty Supernovae Within Wind-Blown Bubbles}
\shortauthors{Mart\'inez-Gonz\'alez  et al.}

\begin{document}

\title{Supernovae within Pre-existing Wind-Blown Bubbles: Dust Injection vs. Ambient Dust Destruction}

\author{Sergio Mart\'inez-Gonz\'alez\altaffilmark{1},
  Richard W\" unsch\altaffilmark{2},
  Sergiy Silich\altaffilmark{3},
  Guillermo Tenorio-Tagle\altaffilmark{3},
  Jan Palou\v s\altaffilmark{2},
  Andrea Ferrara \altaffilmark{4}  
}

\altaffiltext{1}{CONACYT-Instituto Nacional de Astrof\'\i sica, \'Optica y Electr\'onica, AP 51, 72000 Puebla, M\'exico: sergiomtz@inaoep.mx} 
\altaffiltext{2}{Astronomical Institute, Czech Academy of Sciences,  Bo\v{c}n\'\i\ II 1401/1, 141 00 Praha 4, Czech Republic}
\altaffiltext{3}{Instituto Nacional de Astrof\'\i sica, \'Optica y Electr\'onica, AP 51, 72000 Puebla, M\'exico} 
\altaffiltext{4}{Scuola Normale Superiore, Piazza dei Cavalieri 7, I-56126 Pisa, Italy}

\begin{abstract}
By means of 3-D hydrodynamical simulations, here we evaluate the impact that supernova explosions occurring within wind-driven bubbles have on the survival 
or destruction of dust grains. We consider both, the dust generated within the ejecta and the dust initially present in the ambient gas and later locked-up 
in the surrounding wind-driven shell. The collision of the supernova blast wave with the wind-driven shell leads to a transmitted shock that moves into the shell 
and a reflected shock into the ejecta. The transmitted shock is capable of destroying large amounts of the dust locked in the shell, but only if the mass 
of the wind-driven shell is small, less than a few tens the ejected mass. Conversely, massive wind-driven shells, with several times the ejected mass, 
lead upon the interaction to strong radiative cooling, which inhibits the Sedov-Taylor phase and weakens the transmitted shock, making it unable to traverse 
the wind-driven shell. In such a case, the destruction/disruption of the ambient dust is largely inhibited. On the other hand, the SNRs grow 
rapidly in the very tenuous region excavated by the stellar winds, and thus a large fraction of the dust generated within the ejecta is not 
efficiently destroyed by the supernova reverse shock, nor by the reflected shock. Our calculations favor a scenario in which core-collapse supernovae 
within sufficiently massive wind-driven shells supply more dust to the ISM than what they are able to destroy.
\end{abstract}


\section{Introduction}

The propagation of supernova blast waves through their surrounding medium is hold responsible for inducing the disruption (via grain shattering) and destruction 
(via thermal and kinetic sputtering) of a large mass of swept-up interstellar dust \citep[][]{Jonesetal1996,Slavinetal2015}. On the other hand, while core-collapse 
supernovae (SNe) are recognized as efficient dust producers \citep[e.g.][]{TodiniandFerrara2001,Indebetouwetal2014,Matsuuraetal2014}, several authors have argued 
that a large fraction of their ejecta dust will be returned to the gas phase during the thermalization of the SN 
ejecta \citep[e.g.][]{Nozawaetal2007,Micelottaetal2016,Bocchioetal2016,MartinezGonzalezetal2016,MartinezGonzalezetal2017}. It is therefore natural to 
ask whether SNe are ultimately net dust producers or destroyers and under which conditions the answer could be one or the other. 

Aiming to answer this question, \citet{Lakicevicetal2015}, have asserted that supernovae in the Large Magellanic Cloud (LMC) have sputtered more 
ambient dust than what they were able to produce. Their conclusion is based on the analysis of far-infrared and sub-millimeter dust temperature and dust mass 
maps in and around several supernova remnants (SNRs). They, however, were not able to ensure if the dust grains were mostly destroyed or displaced and piled-up. A similar conclusion was drawn by \citet{Temimetal2015}, who inferred the amount of dust sputtered by individual SN blast waves and the 
global rate of grain destruction in the Magellanic Clouds using observationally derived values of the ambient gas density and dust-to-gas mass ratio around 
individual SNRs. However, prior to their explosion, massive stars produce vigorous stellar winds and the role of a pre-existing wind-blown bubble on the 
survival/destruction of interstellar grains has not been considered with sufficient detail. 

The stellar wind produced by a massive star (or a collection of them) piles-up the surrounding ambient gas into a thin, quickly cooling,
expanding shell. The medium surrounding the massive star is then structured (from the center outwards) with a free-wind region, a shocked-wind 
region separated from a wind-driven shell (WDS) by a contact discontinuity and the surrounding undisturbed ambient medium. 

As the massive star explodes, and upon the collision of the SN blast wave with the encompassing wind-driven shell, a reflected 
and a transmitted shocks are generated \citep{TenorioTagleetal1990,TenorioTagleetal1991,Francoetal1991,Rozyczkaetal1993}. 
The crucial factor determining the strength of the shocks is the ratio between the amount of mass collected by the WDS and the mass of the SN ejecta 
($\chi=M_{{wds}}/M_{ej}$). For wind-driven shells with $\chi \lesssim 40$\footnote[1]{The limiting value of $\chi \approx 40$, which determines if the 
SN blast wave is (un)able to overrun an WDS, was calculated for a $10^{51}$ erg explosion with ejecta mass equal to $4$ M$_{\odot}$.}, the SN blast wave rams through and further compresses the WDS and continues to sweep-up the unperturbed 
interstellar medium. On the contrary, if $\chi \gtrsim 40$, the transmitted shock is unable to overrun the WDS and the ambient medium ahead of it remains largely 
undisturbed. These results were later confirmed by \citet{Dwarkadas2005,Dwarkadas2007} and by 
\citet{vanMarleetal2015} and \citet{Haidetal2016}.

\citet{TenorioTagleetal1990} also found that the SNR-WDS interaction is expected to trigger an order of magnitude rise in the X-ray emission. In fact, LMC's 
supernova remnants N63A, N132D, and N49B have been claimed to have exploded within a wind-blown bubble given their X-ray appearance \citep{Hughesetal1998}.

Here we focus on single massive stars that are immersed into wind-driven bubbles prior to their final core-collapse. By performing three-dimensional hydrodynamical 
simulations, we model the collision of an SN blast wave with the pre-existing WDS and then determine the amount of pre-existing ambient dust 
that is destroyed during the pre-supernova wind-driven bubble expansion and after the development of the reflected and transmitted shocks. As 
the crucial parameter is the mass ratio between the WDS and the SN ejecta we fix the parameters related to the stellar wind and the supernova 
explosion, while varying the mass of the wind-driven shell and the density of the ambient medium.

The paper is organized as follows: Section \ref{sec:model} describes our physical and computational scheme, the initial setup (subsection \ref{subsec:ambient}), the stellar wind properties (subsection \ref{subsec:wind}),
and the supernova properties (subsection \ref{subsec:supernova}). In Section \ref{sec:uniform} we discuss the case of supernovae evolving in uniform, homogeneous media,
while Section \ref{sec:bubble} focuses on the more realistic case of supernovae occurring within an encompassing wind-driven shell. In Section \ref{sec:conclusions} we 
outline our major conclusions.

\section{Model Setup}
\label{sec:model}

We have run a set of three-dimensional hydrodynamical simulations with the adaptive mesh refinement code FLASH v4.3 \citep{Fryxelletal2000} to explore 
the explosion of individual massive stars in homogeneous media and within wind-blown bubbles. The hydrodynamic equations are solved with a modified version of the 
Piecewise Parabolic Method \citep[][]{ColellaandWoodward1984}  and the scheme takes into account the equilibrium cooling function of optically 
thin plasmas \citep{Schureetal2009} and the cooling induced by gas-grain collisions 
\citep[calculated using the \textsc{Cinder} module][hereafter MWP+18]{MartinezGonzalezetal2018}. 
With \textsc{Cinder} we also calculate on-the-fly the rate of thermal sputtering given the initial 
distribution of grain sizes and dust mass. Our scheme considers the action of thermal sputtering during the 
whole WDS and the subsequent SNR evolution and assumes a tight coupling between gas and dust. We have generated 
random initial density perturbations (white noise) in order to emulate a degree of clumpiness in the SN ejecta. All the simulations were performed in 
a uniform grid ($256^3$ and $512^3$, as specified later). 

\subsection{Ambient Medium Properties}
\label{subsec:ambient}

The simulations are initialized in a dusty medium with a constant gas number density, $n_{a}$, gas temperature, $T_{a}$, and dust-to-gas 
mass ratio, $\mathcal{D}_{a}$. A gas with one helium atom per every ten hydrogen atoms was considered in all the simulations corresponding to a mean mass 
per particle of ionized and neutral gas $\mu_i = \frac{14}{23} m_H$ and $\mu_n = \frac{14}{11} m_H$, respectively, where $m_H$ is the proton mass.

Since AGB stars and Type II-P SNe \cite[which are the majority of core-collapse SNe][]{Sukhboldetal2016}, tend to form preferentially large 
(and long-lived) dust grains \citep{Asanoetal2013,TodiniandFerrara2001,Nozawaetal2003,Kozasaetal2009}, the ambient grain population is chosen 
to follow a distribution of the form $\sim a^{-1} \exp \{-0.5 [\log(a/a_{0})/\sigma]^2\}$, with $a_{0}=0.1$ $\mu$m and $\sigma=0.7$ and lower 
and upper limits $a_{min}=0.01$ $\mu$m and $a_{max}=0.5$ $\mu$m, respectively, with an equal fraction of silicate and carbonaceous grains. 
The grain mass densities, $\rho_{gr}$, are taken as $2.26$ g cm$^{-3}$ and $3.3$ g cm$^{-3}$ for silicate and carbonaceous grains, respectively.
The grain size distribution is sampled using 10 logarithmically-spaced bins \citepalias[see also Appendix A in][]{MartinezGonzalezetal2018}.

\subsection{Stellar Wind and Wind-Driven Shell Properties}
\label{subsec:wind}

\begin{figure*}
\includegraphics[width=\linewidth]{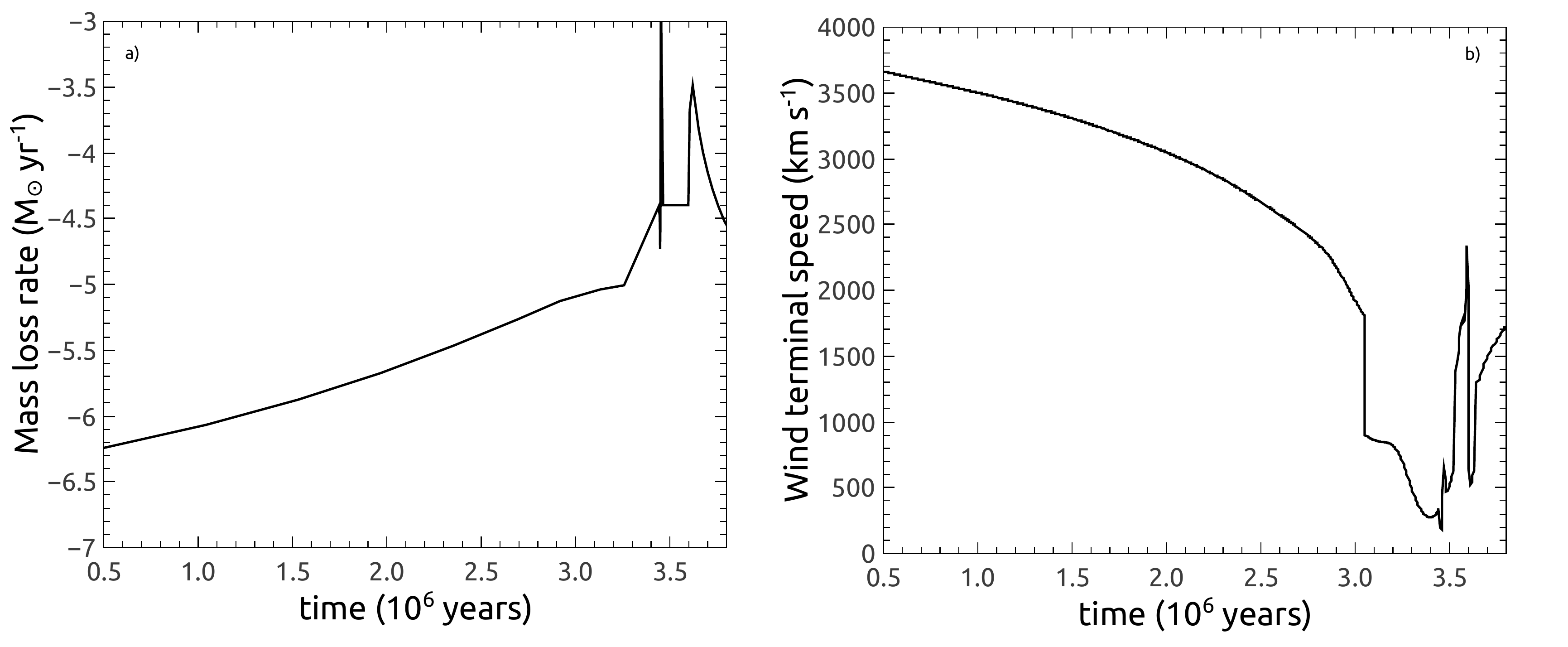}
\caption{Panel {\it a} shows the evolution of the mass loss rate for a 60 M$_{\odot}$ star with solar metallicity obtained by 
\citet{Schalleretal1992}. Panel {\it b} presents the corresponding evolution of terminal speed assuming the 
conversion factor between $v_{\infty}$ and $v_{esc}$ introduced by \citet{Vinketal2001}.}
\label{fig:1}
\end{figure*}

In order to model an isotropic stellar wind we have used the outputs of the stellar evolutionary models by \citep{Schalleretal1992}, 
which span from the beginning of the hydrogen burning phase to the core-carbon-exhaustion. We use, in particular, 
their time-dependent mass loss rate, $\dot{M}_w(t)$, bolometric luminosity, $L_{bol}(t)$, and effective temperature, $T_{eff}(t)$, for stars 
with a solar composition. Having these values, the effective stellar radius, $R_{eff}(t)$, escape velocity $v_{esc}(t)$ and 
stellar wind terminal speed $v_{\infty}(t)$ can be derived \citep[assuming a conversion factor $v_{\infty}=1.3 v_{esc}$ 
if $T_{eff}< 27000$ K and $v_{\infty}=2.6 v_{esc}$ if $T_{eff}\geq 27000$ K,][]{Vinketal2001,SzecsiandWunsch2019}. Figure \ref{fig:1} shows
the evolution of $\dot{M}_{w}$ and $v_{\infty}$ adopted in our stellar wind model.

We have used a modified version of the implementation of a time-dependent wind source by \citet{Wunschetal2008}. 
This approach inserts the wind into a small sphere with radius $R_v$, where the wind mass flux is:

\begin{equation}
\dot{M}(r) = \dot{M}_w r/R_v .
\end{equation}

The gas mass density and velocity around the source are reseted at each timestep as:

\begin{equation}
\rho(r) = \frac{\dot{M}_w}{(4\pi v_\infty r^2)},
\end{equation}
and
\begin{equation}
v(r) = v_\infty r/R_v .
\end{equation}

where $r$ is the distance of a grid cell to the source center\footnote[2]{$r$ has been corrected 
so that it cannot be smaller than the grid cell size.}. In addition to that, the temperature of the wind is set to 
a constant value $10^4$ K. The radius of the source $R_v$ is a free parameters taken as small as possible\footnote[3]{$R_v$ was selected as 
$0.5$ pc in our simulations which allows to have an approximately spherical bubble with the adopted spatial resolution.}.

We have not modeled photoionization nor included the effects of radiation pressure on the dynamics and 
inner structure of the WDS \citep[e.g.][]{MartinezGonzalezetal2014}. We have also not considered any dust produced in stellar winds; however, under 
certain conditions the outflows of massive stars may produce copious amounts of dust prior to their explosion \citep[see ][]{Kochanek2011}. Particularly, 
in the case of colliding stellar winds in close massive binary systems; in eruptive events in luminous blue variables like $\eta$Car \citep{Gomezetal2010}; and 
in the extremely dense ($\sim 10^{10}$ cm$^{-3}$), post-shock cooling layers resultant from the interaction of an SNR with a very dense and slow stellar 
wind \citep{Smithetal2016}. 

\subsection{Supernova Properties}
\label{subsec:supernova}

As the central massive star explodes, it expels a certain amount of mass, $M_{ej}$, whose kinetic energy is $E_{SN}$. We take 
progenitor-dependent values from \citet{Sukhboldetal2016} and insert those values into a sphere of radius $R_{SN}$\footnote[4]{Similarly to 
the chosen value of $R_v$, $R_{SN}=0.5$ pc is taken in our simulations.}. The ejecta mass is composed of gas and dust. The selected initial 
ejecta gas mass density and velocity radial profiles are \citep{TangChevalier2017}

\begin{eqnarray}
\label{eq:ejecta}
 \rho_{ej}=\frac{(3-n)}{4\pi}\frac{M_{ej}}{R_{SN}^3}\left(\frac{R_{SN}}{r} \right)^n ,
\end{eqnarray}

and 

\begin{eqnarray}
\label{eq:vel}
 v_{ej} = \left( 2\frac{(5-n)}{(3-n)}\frac{E_{SN}}{M_{ej}} \right)^{1/2}\left(\frac{r}{R_{SN}} \right) ,
\end{eqnarray}

where $r$ is a radial distance. We have taken $n=2$ in all our simulations as other values of 
$n$ have been shown not to alter significantly the evolution of the SNR 
\citepalias[see Appendix B in][]{MartinezGonzalezetal2018}. At the time of insertion, the ejecta 
is assumed to be at $10^4$ K. 

\section{Explosions in Uniform, Homogeneous Media}
\label{sec:uniform}

\begin{figure*}
\includegraphics[width=0.99999\linewidth]{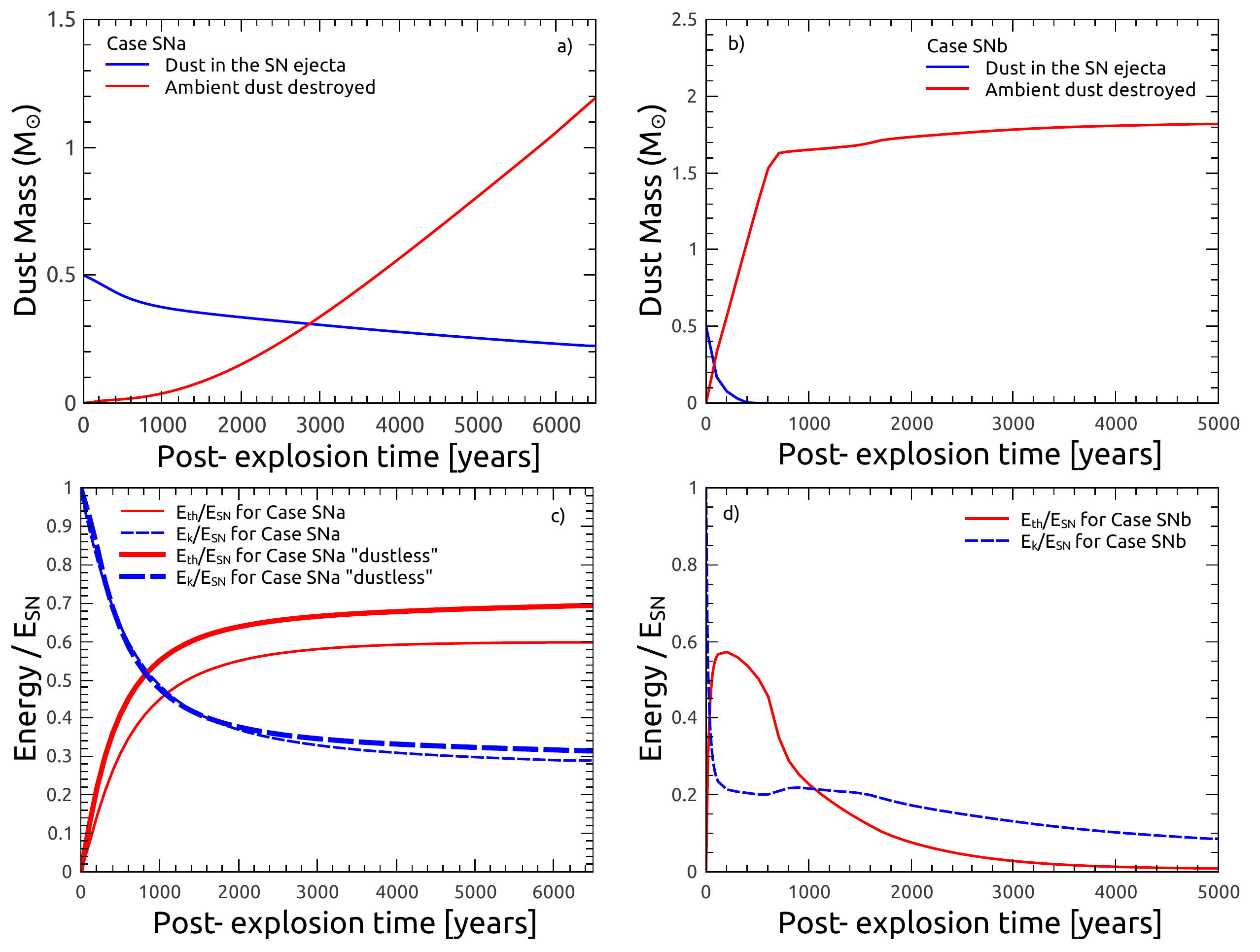}
\caption{Explosions in homogeneous media. The upper panels show the evolution of the SN ejecta dust mass (blue lines), subject 
to reverse shock processing, and the mass of ambient dust destroyed by the SN blast wave (red lines) in 
the case of explosions occurring in homogeneous ambient media (case SNa, panel {\it a}, and SNb, panel {\it b}, 
see Table \ref{tab:1}). The lower panels present the fraction of kinetic (thin dashed blue lines) and thermal 
(thin solid red lines) energies cases SNa (panel {\it c}) and SNb (panel {\it d}). Additionally,
panel {\it c} also displays these fractions for the ``dustless'' case using thicker lines.}
\label{fig:2}
\end{figure*}

\begin{deluxetable}{lcccc}
\tablecolumns{5} \tablewidth{0pc}
\tablecaption{\label{tab:1} \sc Summary of Results}
\tablehead{
\colhead{Model} & \colhead{$n_{a}$} & $\chi$ & M$_{d}^{\rm ej}$ & M$_{d}^{a}$ \\
 & \colhead{\tiny (cm$^{-3}$)} & \colhead{-} & \colhead{\tiny M$_{\odot}$} & \colhead{\tiny M$_{\odot}$}
}
\startdata
SNa  &    $1$ & $0$  & $>0.34$ & $\gg 1.2$ \\
SNa ``dustless'' & $1$ & $0$  & $-$ & $-$ \\
SNb  & $10^3$ & $0$  &  $0.5$ & $2.35$ \\
WDBa  &    $1$ & $400$ & $0.02$ & $\sim 0.45$ \\
WDBb  & $10^3$ & $2\times10^4$ & $0.025$ & $0.28$ 
\enddata
\tablecomments{The Table presents a summary of our results for each model according to the ambient gas number density and the mass 
ratio between the mass of the WDS and the SN ejecta at the time the massive star goes off. M$_{d}^{\rm ej}$ and M$_{d}^{a}$ stand for
the mass of ejecta and ambient dust, respectively, destroyed in each case.}
\end{deluxetable}

We have considered a set of cases with and without pre-existing wind-driven bubbles (WDB). 
In this Section we focus on the latter cases considering SN explosions occurring in ambient media 
with constant temperatures ($T_{a}=10$ K), dust-to-gas mass ratios ($\mathcal{D}_{a}=0.01$) and gas number densities, 
$n_{a}$. The homogeneous low-density case (SNa) has $n_{a}=1$ cm$^{-3}$ and the homogeneous high-density case (SNb) 
assumes $n_{a}=1000$ cm$^{-3}$. For these cases we selected a $60$ M$_\odot$ massive star which expels 
$M_{ej}=5.58$ M$_{\odot}$ ($5.08$ M$_{\odot}$ of 
gas and $0.5$ M$_{\odot}$ of dust) and $E_{SN}=9.12\times10^{50}$ erg when it explodes as a SN. The size of the 
computational domain was selected to be ($20$ pc)$^3$ and ($10$ pc)$^3$ for the SNa and SNb, respectively, and
both in a grid $512^3$. 
Figure \ref{fig:2} shows the evolution of the SN ejecta dust mass and the mass of the ambient dust that is 
destroyed by the SN blast wave in both cases (panel {\it a} and panel {\it b}, respectively). In case SNa the 
SN reverse shock takes $\sim5300$ years to propagate through the whole SN ejecta. At this time: 
$0.25$ M$_{\odot}$ of ejecta dust and $0.88$ M$_{\odot}$ of ambient dust have been destroyed. The reverse shock 
bounces back upon arriving to the SNR's center and subsequently catches up and merges with the blast wave 
\citep{TenorioTagleetal1990,MartinezGonzalezetal2018}. The diameter of the SNR grows to the size of the 
computational domain after $\sim6100$ years when $1.2$ M$_\odot$ and $0.16$ M$_\odot$ of ambient and ejecta 
dust, respectively, have been destroyed. 

For the second model (SNb, see panel {\it b} in Figure \ref{fig:2}), the whole amount of dust 
injected by the SN is destroyed within the reverse shock crossing time ($\sim 500$ years). 
Owing to the higher frequency of ion-grain collisions in the 
dense shell of swept-up ambient material, $\sim1.6$ M$_{\odot}$ of ambient dust are destroyed 
within $\sim 750$ years after the explosion. At this time, radiative cooling (aided by that induced by 
gas-grain collisions) in the shell of swept-up matter becomes catastrophic (the gas temperature drops 
drastically to $\lesssim 10^4$ K), and nearly terminates thermal sputtering. In total, $2.35$ M$_{\odot}$ of 
(ambient+ejecta) dust were destroyed.

In these cases, even without accounting for other destructive/disruptive processes, dust destruction easily overtakes dust production as
suggested by \citet{Lakicevicetal2015}, \citet{Temimetal2015} and \citet{Slavinetal2015}. Additionally, in both cases the inclusion of 
dust-induced radiative cooling provokes a noticeable departure from the classical Sedov-Taylor (ST) solution as it dominates over the 
optically thin radiative cooling of shocked plasmas at temperatures $\gtrsim 3\times 10^{5}$ K. Therefore, the ratios of kinetic energy 
and thermal energy to $E_{SN}$ in case SNa reach only $E_{k}/E_{SN} \approx 0.25$ and $E_{th}/E_{SN}\approx0.6$, respectively, once the 
whole ejecta is thermalized\footnote[4]{In the classical adiabatic ST solution \citep{Sedov1959} 
these values reach $E_{k}/E_{SN} \approx 0.3$ and $E_{th}/E_{SN}\approx0.7$, respectively.}, 
(see panel {\it c} in Figure \ref{fig:2}). Case SNb, which cools catastrophically and ceases dust destruction rapidly, shows a larger 
departure from the ST solution (panel {\it d} in Figure \ref{fig:2}) as $\sim 90\%$ of $E_{SN}$ is radiated away within a few thousand 
years. A ``dustless'' adiabatic case, similar to case SNa but with $\mathcal{D}_{a}=0$ and a dust-free 
SN ejecta was run and we obtained a good agreement with the adiabatic ST solution $E_{k}/E_{SN}\sim 0.32$ and $E_{th}/E_{SN}\sim 0.68$ once the whole SN ejecta is thermalized.

\section{Explosions within Wind-Driven Bubbles}
\label{sec:bubble}

\begin{figure*}
\includegraphics[width=0.91\linewidth]{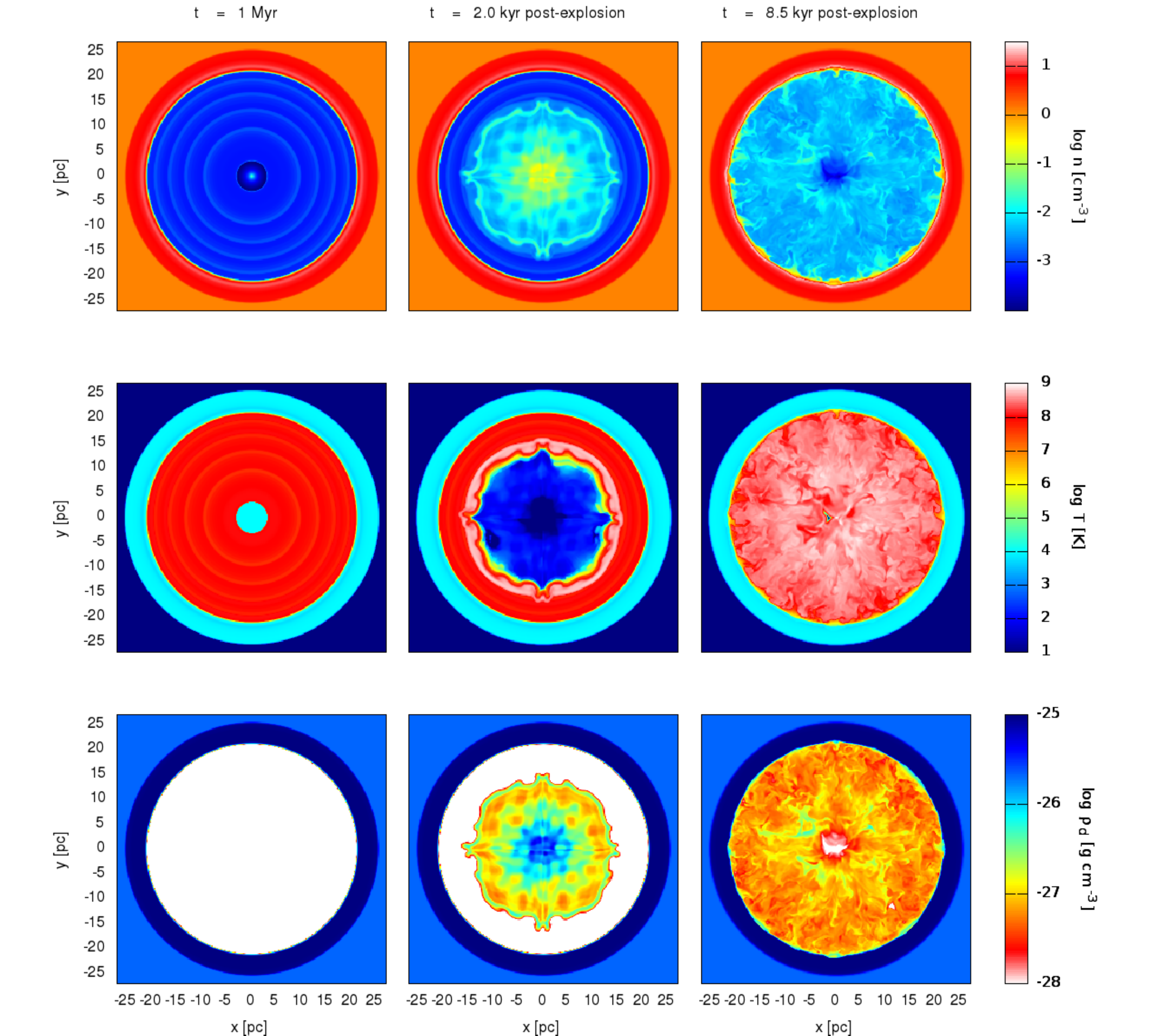}
\caption{
SN explosion within a wind-driven bubble in case WDBa. Two-dimensional cuts along the x-y plane ($z=0$) of 
the distribution of gas number density (upper panels), gas temperature (middle panels), and dust mass density, 
$\rho_{d}$, (bottom panels) at $1$ Myr on the evolution of the stellar wind (left panels), and at 
$2000$ (middle panels) and $8500$ years after the explosion (right panels). After the SNR-WDS collision,
the SN ejecta recoils and fills the wind/SN remnant and the gas density and temperature start to even out. The dust mass density in 
the wind/SN remnant drops mostly because of the SNR expansion rather than due to grain destruction.}
\label{fig:3}
\end{figure*}

\begin{figure*}
\includegraphics[width=0.91\linewidth]{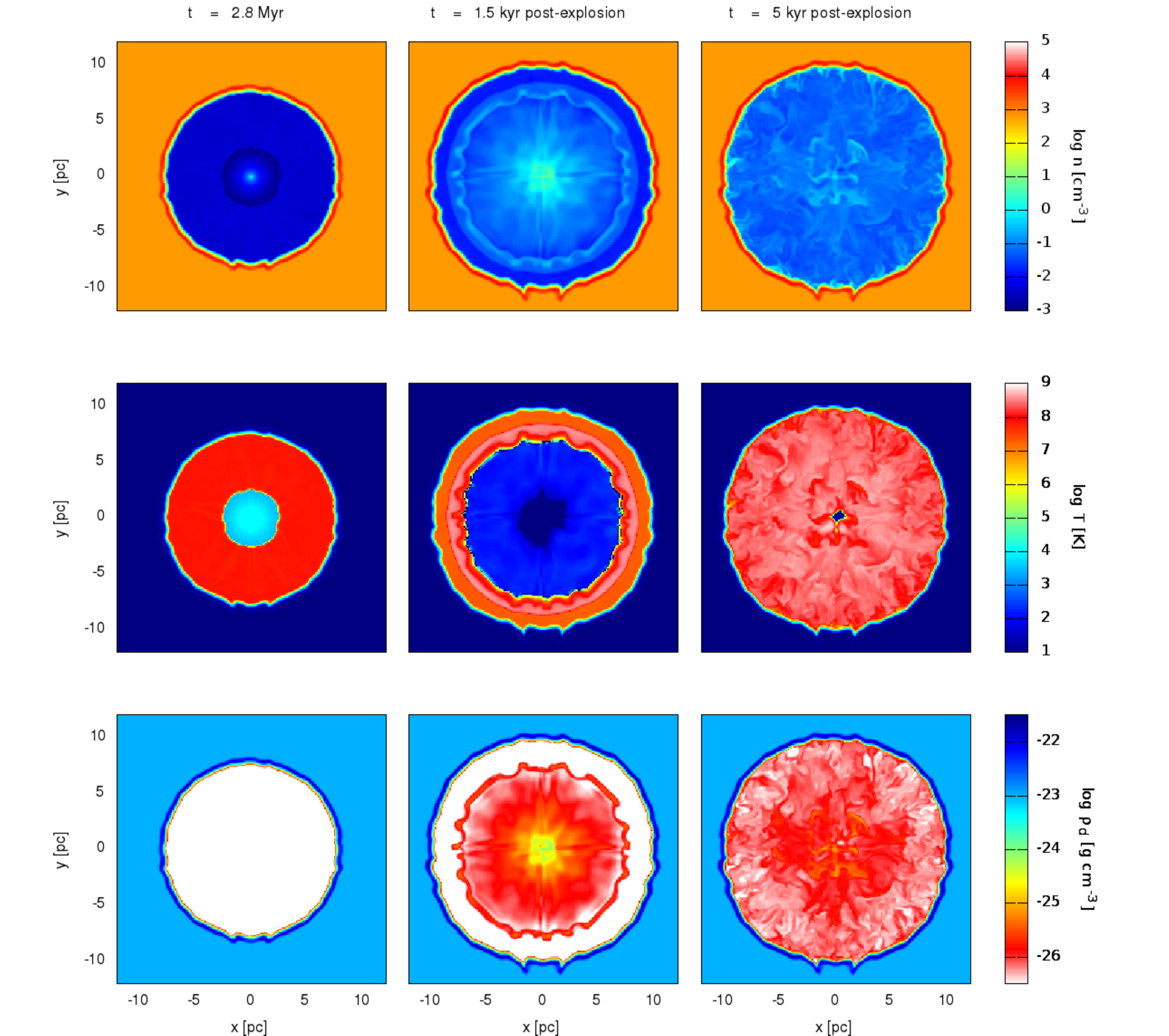}
\caption{Same as Figure \ref{fig:3} but for the WDBb case at $2.8$ Myr on the evolution of the stellar wind 
(left panels), and at 1500 years (middle panels) and 5000 years after the explosion (right panels).}
\label{fig:4}
\end{figure*}

\begin{figure}
\includegraphics[width=\linewidth]{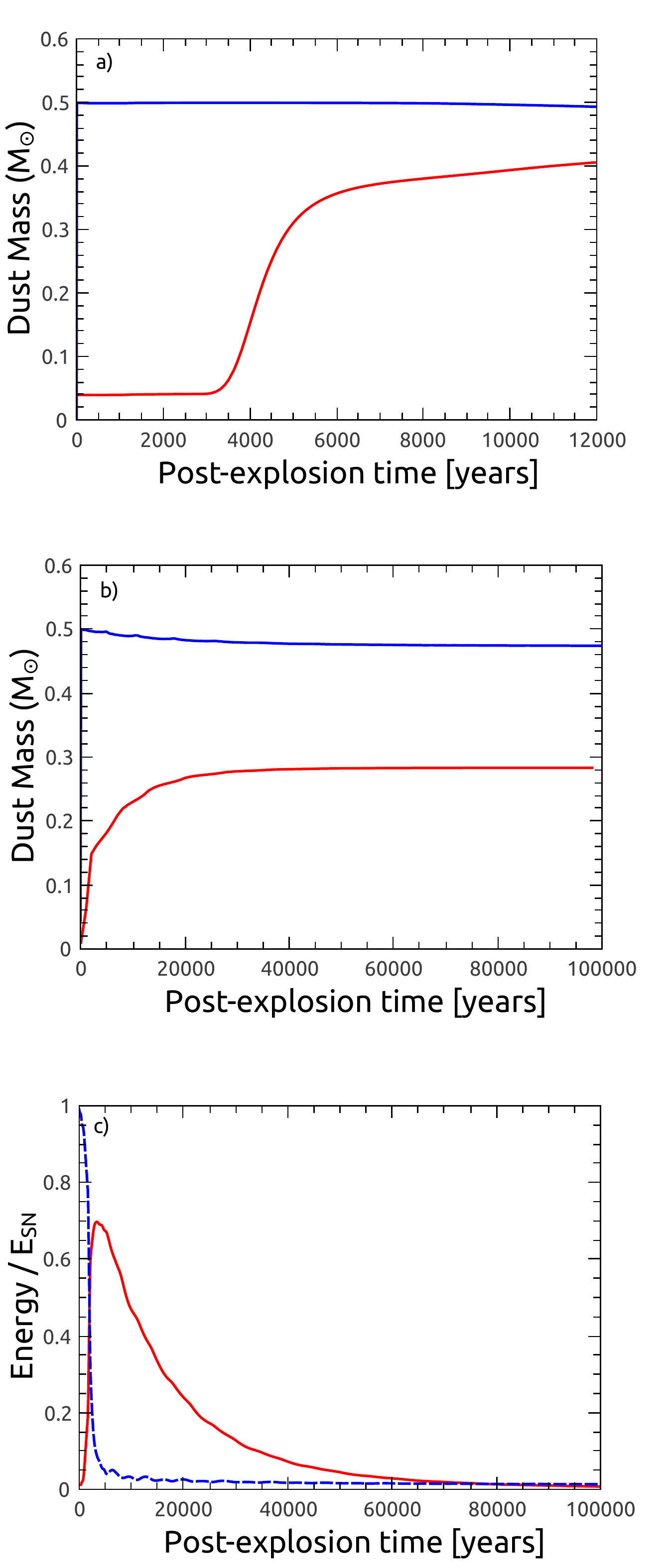}
\caption{Explosions within wind-driven bubbles. Panels {\it a} and {\it b} depict the evolution of the SN ejecta dust mass (blue lines) and the mass of 
destroyed ambient dust by the leading and the transmitted SN blast wave (red lines) in the case of explosions 
occurring within wind-blown bubbles. Panel {\it a} shows the corresponding lines for the WDBa and panel {\it b} displays those of the 
WDBb case (see Table \ref{tab:1}). Panel {\it c} presents the fraction of kinetic (blue dashed lines) and thermal (red solid lines) 
energies for the SNR in case WDBb (see also panel {\it d} in Figure \ref{fig:2})}
\label{fig:5}
\end{figure}

We now focus on the explosion of $60$ M$_{\odot}$ massive stars occurring within wind-blown bubbles.
This choice is justified as SN metal and dust enrichment is particularly important at early cosmic times, before evolved stars 
start to contribute, and when the IMF is thought to be top-heavy \citep[e.g.][]{Schneideretal2002}. Nevertheless, we do not expect a 
qualitatively large difference when studying other progenitor masses given that the crucial parameter which determines the SNR evolution, 
as found by \citet{TenorioTagleetal1990}, is the ratio of the WDS mass to the ejected mass. Indeed, \citet{vanMarleetal2015}, 
\citet{Dwarkadas2007} and \citet{Haidetal2016} explored SNR-WDS collisions in the case of $40$ M$_\odot$, $30$ M$_\odot$ and 
$20$ M$_\odot$ progenitor stars, respectively, and confirmed the evolutionary trends found by \citet{TenorioTagleetal1990}. 

For the purpose of studying bubbles evolving in low-density ambient media ($n_{a}=1$ cm$^{-3}$), 
and in order to maintain a sufficient spatial resolution, we have defined (see Table \ref{tab:1}) 
a ``low-mass WDS'' case (WDBa) in which the central massive star explodes at an arbitrarily 
short time ($1$ Myr), i.e. a shorter time than the predicted stellar evolution for a 60 M$_\odot$. The mass of the 
wind-driven shell implies a value of $\chi \approx400$. In this case, only $0.04$ M$_{\odot}$ of the dust present in the WDS is 
destroyed prior to the supernova explosion (see Appendix \ref{app:A}).

Before the collision of the SNR-WDS, and as the SNRs evolve in the tenuous region excavated by the stellar 
winds, only a small fraction of the SN ejecta is thermalized by the reverse shock and about 
$\sim 4\%$ ($\sim 0.02$ M$_{\odot}$) of the ejecta dust is destroyed in case WDBa.

In this case, the SNR-WDS collisions occurs about $\sim 3200$ years after the SN explosion. The left panels in Figure 
\ref{fig:3}, show the distribution of the gas number density, temperature and dust mass density in the wind-blown bubble at 
a stellar evolutionary time $1$ Myr for the WDBa case. Prior to the SN explosion, the wind-driven bubble shows its 
four-zone structure and one can note the piling-up of the gas and dust in the WDS. The middle panels show the same quantities 
but a two thousand years after the SN explosion. At this time, the SNR is sweeping-up the wind matter and the blast wave is 
approaching to collide with the WDS. When the SNR-WDS collision occurs, the procuded reflected shock catches-up and merges with 
the SN reverse shock, reaches the remnant's center and then transforms into a subsonic forward wave.

As illustrated in the right panels of Figure \ref{fig:3} for the WDBa case, the SN blast wave is unable 
to overrun the wind-driven shell and the SNR ends-up being confined to roughly the size of 
that the wind-driven bubble had reached at $1$ Myr. Panel {\it a} in Figure \ref{fig:5} shows 
that the wind-driven shell, let to evolve only during $1$ Myr, is massive enough to prevent the blast wave
to destroy a larger mass of ambient dust than the amount of dust able to survive in the SN ejecta.

Limited spatial resolution inhibits a complete calculation of a low-density case in which we could follow 
the full evolution of the wind-driven bubble during $3.8$ Myr. This is due to the very large radius that 
the bubble would acquire that would prevent us to sufficiently resolve the early stages of the SNR evolution. However, it 
can be envisaged that an even smaller amount of ambient dust would be destroyed if the blast wave encounters an even more 
massive WDS than in case WDBa (see Table \ref{tab:1}).

We have also studied a high-density case (WDBb, $n_{a}=1000$ cm$^{-3}$) in which we let the massive star inject 
its stellar wind until core-carbon-exhaustion occurs given that the final core-collapse will proceed shortly 
after ($\sim 3.8$ Myr). This simulation is inscribed into a cube ($54$ pc)$^3$ in a grid $256^3$.
 
Upon the SNR-WDS collision, the transmitted shock moves initially at a velocity of a few 
$\sim 1000$ km s$^{-1}$. However, at $3.8$ Myr the massive WDS is four orders of magnitude more massive than 
the SN ejecta and thus this velocity cannot be sustained for a long distance and sharply drops 
\citep{Dwarkadas2007}. This also limits the relative importance of other grain disruption mechanisms which require
high-velocity shock processing (e.g. kinetic sputtering and grain shattering, see Appendix \ref{app:B}). 
 
As depicted in panel {\it a} in Figure \ref{fig:5}, only $\sim 5\%$ ($0.025$ M$_{\odot}$) of the dust mass 
injected by the SN are destroyed. Not only that, but about $\sim 0.28$ M$_{\odot}$ of ambient dust are 
destroyed ($\sim 0.025\%$ of the total amount of swept-up ambient dust). Thus, in both WDBa and WBDb cases, the massive WDS 
poses an almost unsurmountable barrier that prevents the SN blast wave from processing the majority of the 
ambient dust locked in the WDS. 

In both WDB cases, the Sedov-Taylor phase is totally inhibited by strong radiative cooling, which becomes dominant 
early during the SNR evolution. Therefore, the energy of the SN explosion is quickly radiated away rather than used to 
sustain gas and dust thermal collisions (see panel {\it c} in Figure \ref{fig:5} for the WDBb case). 

The survival of ejecta dust is favored because of the SNR expansion within the low-density medium excavated by 
the stellar wind, similarly to what was found by \citetalias{MartinezGonzalezetal2018} for the case of 
clustered SNe evolving in a collective star cluster wind. On top of that, kinetic sputtering and 
grain shattering (not included in our simulations) will most likely be disfavored given that they also
require a sufficient rate of gas-grain and grain-grain encounters and not only a large relative motion
between impinging particles/grains (see Appendix \ref{app:B}). As a result, the consideration of the full stellar 
mass-loss history has a profound impact on the survival of both, the ambient and the ejecta dust grains.

Note also that the presence of low dust density regions around SNRs does not necessarily imply, as previously suggested, that 
core-collapse SNe destroy more dust than what they produce (see bottom panels in Figure \ref{fig:4}). Moreover, the SN ejecta dust
will mostly radiate at near- to mid-infrared wavelengths, and therefore will be more difficult to observe in far-infrared and 
sub-millimeter maps.

We have neglected the presence of interstellar magnetic fields which can increase the 
thickness of the WDS as it re-expands driven by magnetic pressure, thus decreasing the mass density of the WDS 
\citep{Ferriereetal1991}. In that scenario, the transmitted shock could propagate farther into the WDS than in the absence 
of a magnetic field. However, not only the gas but also the dust mass density is reduced and thus the timescale for 
grain destruction within the WDS is increased ($\tau_{dest}\sim n^{-1}$) \citep[see also ][who showed that even in the presence of 
a strong interstellar magnetic field, an SN blast wave is still unable to overrun its encompassing massive WDS]{vanMarleetal2015}.
Thus, the presence of an insterstellar magnetic field is unlikely to change our conclusions significantly.

\section{Concluding Remarks}
\label{sec:conclusions}

Previous studies have estimated that SN explosions in homogeneous ambient media destroy more dust than what SNe are able to inject 
\citep[e.g.][]{Slavinetal2015,Lakicevicetal2015,Temimetal2015}. However, these estimates did not take into consideration the shaping of the 
interstellar gas during the pre-SN massive star evolution. During this stage, powerful stellar winds evacuate the ambient gas from 
the stellar vicinity, compressing it into an expanding dense shell. This may affect the supernova remnant evolution significantly. 

Indeed, as shown by \citet{TenorioTagleetal1990} and \citet{Francoetal1991}, an SN blast wave is able to overrun an encompassing wind-driven 
shell only if its mass, $M_{wds}$, is small and comparable to mass of the SN ejecta, $M_{ej}$. In the other case, when $M_{wds}\gg M_{ej}$, 
the WDS becomes an unsurmountable barrier that confines the SNR within the wind-driven bubble.

Here we have shown, by means of three-dimensional hydrodynamical simulations, that this also affects the fate of pre-existing ambient and SN ejecta 
dust grains. The ambient dust grains accumulated in a massive wind-driven shell remain largely unaffected by thermal sputtering after the 
SNR-WDS collision. The dust ejected by SNe is also mostly unaffected as the SNRs expand rapidly in the tenuous region previously 
excavated by the pre-existing stellar wind. Destruction of ejecta dust in such cases is caused, rather than by the passage of the reverse 
shock alone, by the reflection of the blast wave that catches-up with the reverse shock. In these cases, radiative cooling proceeds very 
rapidly and the SNRs do not pass through the Sedov-Taylor phase. Other grain destructive/disruptive processes are expected to be also 
inhibited in the WDS as the transmitted shock is weak and penetrates only into a very thin inner layer of the WDS. This also prevents
the efficient mixing between the WDS matter and that from the wind/SN remnant.

This situation is radically different from that occurring when the explosion is modeled in a homogeneous medium or embedded in a low-mass 
wind-driven shell, where one can expect that the blast wave and the reverse shock could destroy, by far, more dust than what can be produced 
after the SN explosion.  

This study, together with the three-dimensional hydrodynamical simulations for the case of clustered SNe presented by 
\citet[][]{MartinezGonzalezetal2018}, show that core-collapse SNe may supply dust efficiently to the ambient medium and that they do not 
destroy large amounts of the pre-existing dust in the surroundings. As the fraction of surviving ejecta dust might be very high, this result 
is also consistent with recent estimates of little dust destruction by SNe in the Local Group and in high-redshift galaxies 
\citep{GallandHjorth2018,Michalowskietal2019} and theoretical expectations that suggest that the bulk of dust in the early Universe must come 
from core-collapse SNe \citep{Ferraraetal2016}.

The authors thankfully acknowledge the computer resources, technical expertise and support provided by the 
Laboratorio Nacional de Superc\'omputo del Sureste de M\'exico, CONACYT member of the network of national 
laboratories. The authors thank the anonymous Referee for a careful reading and helpful suggestions which 
greatly improved the paper. This study was supported by CONACYT-M\'exico research grant A1-S-28458. S.M.G. also acknowledges support by CONACYT 
through C\'atedra n.482. R.W. and J.P. acknowledge the support from project 19-15008S of the Czech Science Foundation and from the 
institutional project RVO:67985815 and the support by The Ministry of Education, Youth and Sports of the Czech Republic from the
Large Infrastructures for Research, Experimental Development and Innovations project "IT4Innovations National Supercomputing Center – LM2015070".
A.F. acknowledges support from the ERC Advanced Grant INTERSTELLAR H2020/740120. Any dissemination of results 
must indicate that it reflects only the author’s view and that the Commission is not responsible for any use that may be made of the information it 
contains. This research was supported by the Munich Institute for Astro- and Particle Physics (MIAPP) of the DFG cluster of excellence ``Origin and 
Structure of the Universe''. Partial support from the Carl Friedrich von Siemens-Forschungspreis der Alexander von Humboldt-Stiftung Research Award 
is kindly acknowledged. S.M.G. expresses gratitude to the Czech people for their hospitality during his 2016-2018 postdoctoral stay and the 
Erasmus+ programme of the European Union under grant number 2017-1-CZ01-KA203-035562 for his 2018 stay at the Instituto de Astrof\'isica de Canarias.

\appendix*
\section{Appendix}
\subsection{Ambient Dust Processing during the Pre-SN Evolution}
\renewcommand{\theequation}{A\arabic{equation}}
\label{app:A}

We have taken into consideration the action of thermal sputtering during the whole pre-SN evolution of 
the wind-driven bubble. However, the ambient dust locked in the WDS is not significantly affected by 
thermal sputtering as the WDS cools down quickly to $\sim 10^4$ K.

This occurs in a timescale, $\tau_{cool}$, approximately given by \citep{MacLowMcCray1988}:

\begin{eqnarray}
      \label{taucool}
\tau_{cool} &=& (2.3 \times 10^4) \left(\frac{Z_{a}}{\mbox{Z}_{\odot}}\right)^{-0.42} \left(\frac{n_{a}}{\mbox{1 cm}^{-3}}\right)^{-0.71} \times \left( \frac{L_{w}}{10^{38} \mbox{ erg} \mbox{ s}^{-1}}\right)^{0.29} \mbox{ years} ,
\end{eqnarray}

where $Z_{a}$ is the metallicity and $n_{a}$ is the number density of the ambient medium and $L_{w}=\frac{1}{2}\dot{M}_{w}v_{\infty}^2$ is the 
wind mechanical luminosity. For cases  and WDBa, the WDS cools down in $\sim 250$ kyr 
leading to the destruction of only $0.01$ M$_{\odot}$ of the dust locked in the WDS. For the WDBb, the WDS 
cools down in less than $\sim 2000$ years. These simple estimates of $\tau_{cool}$ ignore the contribution 
of dust grains to the cooling of the WDS; when taken into account \citep[as in our numerical scheme and that 
by][]{EverettandChurchwell2010} these timescales are greatly reduced and therefore destruction of the ambient 
dust in the WDS is suppressed. Not only that but in the low-density cases, thermal sputtering becomes also 
inefficient as the number density inside the WDS is only a few particles per cm$^{-3}$.

\subsection{The Role of Kinetic Sputtering and Grain Shattering}
\label{app:B}

Our simulations have ignored some processes that can be important for grain destruction, i.e. kinetic 
sputtering and grain shattering as a result of grain-grain collisions. The disruption timescales via kinetic sputtering and 
grain shattering are  \citep[e.g. ][]{HoangandTram2019}:

\begin{eqnarray}
 \tau_{sp}^{k}=\frac{4a\rho_{gr}}{n m_{H} Y_{sp}^ {k} v_{gr}} ,
\end{eqnarray}

and 

\begin{eqnarray}
 \tau_{gg}=\frac{4a\rho_{gr} \mathcal{D}}{3 n m_{H} \delta v} ,
\end{eqnarray}

where $n$ is the gas number density, $Y_{sp}^{k}$ is the sputtering yield, $v_{gr}$ is the 
relative speed between gas and dust grains, $\mathcal{D}$ is the dust-to-gas mass ratio and $\delta v$ is the 
relative speed between colliding dust grains. 

The upper panels in Figures \ref{fig:3} and \ref{fig:4} show that the number density of the bulk of the SN ejecta drops to 
$n\sim 10^{-1}-10^{-2}$ cm$^{-3}$ when it is crossed by the reverse/reflected shock. For characteristic values of 
$a=0.01$ $\mu$m, $Y_{sp}^{k}=0.1$ \citep{HoangandLee2019}, ejecta dust-to-gas mass ratio 
$\mathcal{D}\approx0.5/5.08\sim10^{-1}$, $\rho_{gr}=3$ g cm$^{-3}$, $v_{gr}=175$ km s$^{-1}$ \citep{Fryetal2018} and 
$\delta v_{gr}=20$ km s$^{-1}$ \citep[][]{HoangandTram2019}, the disruption timescale due to kinetic sputtering 
is $\tau_{sp}^{k}\sim 1.4-14$ Myr, and that corresponding to grain-grain collisions is $\tau_{gg}\sim 4-41$ Myr. The 
timescale for grain damping by gas-grain collisions is \citep{Hoangetal2012}:

\begin{eqnarray}
 \tau_{damp}=\frac{a\rho_{gr} }{n}\left(\frac{8 m_{H}k_{B} T}{\pi}\right)^{-1/2} ,
\end{eqnarray}

where $k_{B}$ is the Boltzmann constant and $T\sim 10^8$ K is the gas temperature. $\tau_{damp}$ is 
much shorter ($\sim4-40$ kyr) than $\tau_{ksp}$ and $\tau_{gg}$ and thus these processes are not important for grains immersed into
the shocked ejecta when the SNR evolves within a wind-driven bubble. 

In the case of ambient dust grains locked-up in the WDS, the fact that the velocity of the transmitted 
shock drops sharply and goes quickly below the threshold shock velocity for grain shattering 
\citep[$\sim 40$ km s$^{-1}$,][]{Jonesetal1996} inhibits kinetic, thermal sputtering and grain-grain collisions 
for the vast majority of the WDS.

\bibliographystyle{apj}
\bibliography{Infrared}

\end{document}